\documentclass[aps,12pt,preprint,tightenlines]{revtex4}
\usepackage{color,graphicx}
 \pdfoutput=1 

\begin{document}
\newfont{\bftitle}{cmbx12 scaled\magstep5}
\newcommand{\boldsigma}{\mbox{\boldmath $\sigma$}}
	\def\bfa{{\bf a}}
	\newcommand {\boldnabla}{\mbox{\boldmath$\nabla$}}
\newcommand{\boldgamma}{\mbox{\boldmath $\gamma$}}
\newcommand{\boldepsilon}{\mbox{\boldmath $\epsilon$}}
\def\poinc{Poincar\'{e} }
\def\bfq {{\bf q}}
\def\bfr {{\bf r}}
	\def\bfa{{\bf a}}
\def\bfalpha{{\bf\alpha}}
\def\bfK{{\bf K}}
\def\bfL{{\bf L}}
\def\bfk{{\bf k}}
\def\bfS{{\bf S}}
\def\bfp{{\bf p}}
\def\be{\begin{equation}}
 \def \ee{\end{equation}}
\def\bea{\begin{eqnarray}}
\def\non{\nonumber\\}
  \def\eea{\end{eqnarray}}
\newcommand{\pbar}{\bar{p}_\lambda(g)}
\newcommand{\eq}[1]{Eq.~(\ref{#1})}
\newcommand{\eqn} {Eq.~(\ref )}
\newcommand{\rhol}{\rho_\lambda(g)}
\newcommand{\bb}{\langle}
\newcommand{\alphab}{1-\alpha}
\newcommand{\kk}{\rangle}
\newcommand{\bk}[4]{\bb #1\,#2 \!\mid\! #3\,#4 \kk}
\newcommand{\kb}[4]{\mid\!#1\,#2 \!\mid}
\newcommand{\kx}[2]{\mid\! #1\,#2 \kk}
\newcommand{\tmu}{{\mu}}
\def\notq{{\not\! q}}
\def\notp{{\not\! p}}
\def\notk{{\not\! k}}
\def\up{{\uparrow}}
\def\down{{\downarrow}}
\def\bfb{{\bf b}}
\def\bfP{{\bf P}}
\def\bfE{{\bf E}}
\title{
 Clustering Coefficients  of  Protein-Protein Interaction Networks}

\author{Gerald A. Miller$^*$, Yi Y. Shi$^\dagger$, Hong Qian$^\dagger$, and Karol Bomsztyk$^\ddagger$}
\affiliation{ Departments of $^*$Physics, $^\dagger$Applied Mathematics, and $^\ddagger$Medicine,\\
University of Washington
  Seattle, WA 98195}

\sloppy

\begin{abstract}The properties of certain networks are determined by hidden variables that 
are not explicitly measured. The conditional probability (propagator) that 
a vertex with a given value of the hidden variable is
connected to k of other vertices determines all measurable properties.  
We study  hidden  variable
models and    find an 
averaging
approximation that enables us to obtain a general analytical result for the propagator.
Analytic results showing the validity of the approximation are obtained.
 We apply   hidden variable models
 to protein-protein interaction networks (PINs) in which 
the hidden variable is  the association
free-energy, determined by distributions that depend on biochemistry and evolution.   
We  compute degree distributions 
as well as
clustering coefficients of several PINs of different species; 
good agreement with measured data is
obtained. For the human interactome two 
different parameter sets give the same degree distributions, 
but the computed clustering coefficients differ by a factor of about two. 
This shows that degree distributions are not sufficient to determine the properties
of PINs.
\end{abstract}
\maketitle
\vskip0.5cm

\section{Introduction}
\label{sec:intro}

Physicists have recently shown that network analysis is  a powerful  tool to study
the statistical properties of  complex
biological, technological and  social systems of diverse kinds\cite{str,albert,newmanrev}. 
Many  networks 
exhibit
 a 
scale-free degree distribution in which the probability $p_k$ that a vertex is connected to
$k$ other vertices falls as a power $p_k \sim k^{-\gamma}$. 
This property is not 
sufficient to completely  describe natural networks because such 
systems also  exhibit degree {correlations}--
the degrees of the vertices at the
end points of any given edge are not independent
\cite{alexei,alexei02,nprl,newmanmixing}.  It is not 
surprising that natural systems depend on properties
that
do not appear explicitly in degree distributions. In particular, 
 protein interaction networks 
depend on the availability of  sufficient binding free energy\cite{cell}
 to cause interactions to occur 
(links between vertices
to exist). 

 Caldarelli \textit{et al.} \cite{gcalda03} and S{\"o}derberg
\cite{soderberg}  proposed 
models
in which vertices are  characterized by a 
{fitness} parameter  assigned according to a chosen  probability
distribution. Then, pairs of vertices are independently joined by an
undirected edge with a probability depending on the fitnesses  of the
 end points.
Ref.~\cite{bps} 
generalized these models as a class of models with {hidden variables} and 
presented
a  detailed formalism   showing how 
to compute  network
properties using the  conditional probability (propagator)  
that a vertex with a given value 
of a  hidden variable is 
connected to other $k$ vertices.
 This  formalism, valid for any Markovian (binary) network,  
provides the generating
function for the propagator, but not the  propagator itself.

The   purpose of this paper  is twofold. We first use  a    mean field 
 approximation to derive  a 
  general analytic formula for the propagator, therefore finding a general  approximate
solution to to the inversion problem.
This enables one to compute network properties without the use of a simulation procedure,
thereby simplifying the computational procedure and potentially broadening the ability
of scientists from all fields to use network theory. The validity of the  method
is assessed by comparing the results of using our approximation with published results.  
We then use this method to compute  clustering coefficients
of  a specific hidden variable model  for protein-protein interaction networks (PIN) 
from several organisms developed by us\cite{us} that previously had obtained 
degree distributions  in agreement with  measured data. 
We show  that two models with  the
same degree distribution have  very different clustering coefficients.

We  outline this in more detail. Sect.~\ref{sec:formalism} reviews the hidden
variable formalism and our approximate solution to the inversion problem. We distinguish
 between sparse (which have been solved in Ref.~\cite{bps})
and non-sparse networks  which are solved here. The next section \ref{sec:models}
studies the models of Refs.~\cite{gcalda03} and \cite{deeds}. Our averaging procedure
is found to work well for most situations.  Our own model\cite{us} is presented
in \ref{sec:PIN}. We present an analytic result for the average connection probability
and extend the results of \cite{us} to computing the clustering coefficients. The final
section \ref{sec:summary} is reserved for a brief summary and discussion.

\section{Hidden Variable Networks}
\label{sec:formalism}

We 
present the formalism for  hidden variable models \cite{bps}. 
The probability that  a node has a hidden
continuous variable $g$ is given by $\rho(g)$, normalized so that its 
integral over its domain
 is unity. 
This function is chosen to be an exponential in 
\cite{gcalda03,us} and a Gaussian in 
\cite{deeds}. 
 The connection probability for 
two nodes of $g,g'$ is defined to be $p(g,g')$. 
This is taken as a step function in \cite{gcalda03,deeds}, and a
Fermi function in \cite{us}. 
The two functions $\rho(g)$ and $p(g,g')$ can be chosen in a wide variety of ways
to capture the properties of a given network.
Reference~\cite{bps}
presents the  probability generating function, $G_0(x)$, that determines $p_k$
 in terms of the generating function for the
propagator, $\widehat{G}_0(z,g)$, as
\bea G_0(z)=\int 
dg\rho(g)
\widehat{G}_0(z,g),\label{bog1}\eea where 
\bea
\ln \widehat{G}_0(z,g)=
N\int 
dg' \rho(g')\log(1-(1-z)p(g,g')).  \label{gbog}\eea
The propagator 
$G_0(k,g)$ 
giving the conditional 
probability that a vertex of hidden variable
$g$ is connected to $k$ other vertices is given implicitly by
\bea
\widehat{G}_0(z,g)=\sum_{k=0}^\infty\;z^k\;G_0(k,g).
\label{G0kg}\eea 
Knowledge of  $G_0(k,g)$ determines the conditional probability $P(k'|k)$ that a node
of degree $k$ is connected to a node of  degree $k'$, 
 \cite{bps} (as well as $p_k$), and those two functions completely define a Markovian network.
Once $G_0(k,g)$ is the determined, all of the properties of the given network are determined.
The most well-known example is the degree distribution $p_k$:
\bea p_k=\int_0^\infty dg \rhol G_0(k,g).
\eea

It would seem that determining  $G_0(k,g)$ from \eq{gbog} is a simple technical matter,
but this is not the case\cite{bps}. The purpose of the present Section is to provide
a simple, analytic and accurate method to determine  $G_0(k,g)$. 

We   obtain $G_0(k,g)$ 
from \eq{gbog} by using the tautology
\bea p(g,g')= \bar{p}(g) + (p(g,g')-\bar{p}(g)\label{exp})\eea
 in \eq{gbog}, 
choosing $\bar{p}(g)$  so as to eliminate the effects of the second term,  and then 
treating  the  remaining higher powers of  $(p(g,g')-\bar{p}(g))$
as an expansion parameter. 
Using \eq{exp} in  \eq{gbog} yields
\bea&&
\ln \widehat{G}_0(z,g)=
\ln \widehat{G}_0(z,g)=\log(1-(1-z)\bar{p}(g))^N-
N(1-z)\int dg'\rho(g'){(\bar{p}(g)-p(g,g'))\over 1-(1-z)\bar{p}(g)}\non&&
-N\sum_{n=2}^\infty\; 
{(1-z)^n\over n}\int dg'\rho(g')\left({p(g,g')-\bar{p}(g))\over
1-(1-z)\bar{p}(g)}\right)^n.  \label{gbog01}\eea  
In analogy with  the mean-field (Hartree)
 approximation of atomic and nuclear
physics, we find that the second term of \eq{gbog01} vanishes if we choose
  $\bar{p}(g)$ to be 
 the average of $p(g,g')$ over $\rho(g')$:
\bea
\bar{p}(g)=\int\;dg'\rho(g')p(g,g').\label{pave}\eea
With \eq{pave} the effects of the term of  first order in $(p(g,g')-\bar{p}(g))$  vanish.
We therefore obtain the result:
\bea
\ln \widehat{G}_0(z,g)=\log(1-(1-z)\bar{p}(g))^N-N\sum_{n=2}^\infty\; 
{(1-z)^n\over n}\int dg'\rho(g')\left({p(g,g')-\bar{p}(g))\over
1-(1-z)\bar{p}(g)}\right)^n,  \label{gbog1}\eea  
with the putative term with $n=1$ vanishing by virtue of \eq{pave}.

We treat the first term of \eq{gbog1} as the leading order ($LO$) term and regard the
remainder as a correction.  The validity of this approach can be checked by comparison
with simulations, or (in certain cases) with analytic results. 
 Numerical results for the  PIN of current interest \cite{us} 
indicate that the corrections to the \textit{LO} terms induce errors in 
$p_k$ of no more than a few percent and
that the approximation becomes more accurate for large values of $k$.
 Therefore we use the  \textit{LO} approximation. 
 Using  
exponentiation and the binomial theorem in  the 
first  term of \eq{gbog1} leads to the result
\bea
{G}^{(LO)}_0(k,g)=\left(\begin{array}{c}N\\k\end{array}\right) (1-\bar{p}(g))^{N-k}\bar{p}(g)^k,\label{glo}\eea
which is of the form of a random binomial 
distribution in which the connection probability depends on the
hidden variable $g$. The \eq{glo} is our central new general  result that can be used for any
hidden variable network. This binomial distribution has both the normal Gaussian
and Poisson $(N p(g)\ll1)$  distributions as  limiting cases. 

\subsection{Sparse and Nonsparse Networks}
Ref.~\cite{bps} explained the difference between sparse
 and nonsparse networks. Sparse networks have a well-defined thermodynamic limit for the average degree, while
this quantity diverges as the network size $N$ approaches
infinity. Ref.~\cite{bps} defines criteria for sparseness
by pointing out the relevance
of $\bar{p}$ of \eq{pave} in determining whether or not
a network is sparse. Given this quantity the average
degree is \bea
\langle k \rangle=\int dg \rho(g)\bar{p}(g)=
\int dg\int dg' \rho(g)p(g,g')\rho(g').\eea
If the $\rho(g)$ is independent of $N$ the only way
to obtain a non-divergent value $\langle k \rangle $
is for the connection probability \cite{bps} to scale as $N^{-1}$:
\bea
p^{\rm sparse}(g,g')={C(g,g')\over N},\; {\rm sparse\; network}\;
.\label{sparse}\eea 
Under the specific  assumption that   \eq{sparse} holds,
 Ref.~\cite{bps} finds a very interesting result.   In our
notation, this amounts to 
using \eq{sparse} in \eq{gbog} and taking
the limit that $N$ approaches infinity. Then 
\bea
G_0^{\rm sparse}(z,g)=\exp(z-1)\int dg'\rho(g')C(g,g').
\eea
This 
shows that the Poisson limit of \eq{glo} is obtained for the very 
special case of sparse networks in which  the connection probability scales as $N^{-1}$.
None of the models of interest here \cite{gcalda03,us,deeds} are sparse, 
so it is our  present result (\ref{glo})
that is  widely applicable.

\subsection{General Networks}
Turning to the use of the use of the propagator, we obtain  
the 
degree distribution as \bea p_k=\int dg \rho(g) G_0(k,g)\approx
\int dg \rho(g) G_0^{(LO)}(k,g).
\label{ours}\eea
This expression can be thought of as averaging
a  binomial distribution over the hidden variable and is a natural 
generalization of
classical graph theory.
A  similar expression for $p_k$ has been obtained, in the Poisson limit,
in Ref.~\cite{thurner}. In that work, $p_k$ 
is presented as an 
integral of the Poisson distribution for $p(g)$ multiplied by
 the ``$P$ representation'' of a density matrix. Comparing \eq{glo} with the result (3) of
\cite{thurner} shows that our propagator is proportional to  the $P$ representation, 
 essentially our $\rho(g)$. 
Ref.~\cite{thurner} shows, how under certain assumptions, to use
$p(k)$ 
to determine the
$P$ representation.  Our method allows 
underlying
network properties, denoted by $\rho(g)$ and $p(g,g')$, to predict various
network properties.

The clustering 
 coefficient which measures  transitivity \cite{newmanrev}: 
if vertex $A$ is connected to vertex $B$ and vertex $B$ to vertex $C$, there is an
increased probability that vertices  $A$ and $C$ are connected.   In graph theory, 
the clustering coefficient
$c(k)$ is the ratio of the number of triangles to the number of pairs, computed for
nodes of degree $k$.    Ref.~\cite{bps} shows that 
\bea
&&c(k)=\frac{1}{p_k}  \int
dg  \rho(g)G_0(k,g)c(g) \label{eq:21}\\&&
c(g)= \int 
dg'  \int 
dg''{\rho(g')p(g,g')\over\bar{p}(g)} p(g',  g''){\rho(g'')p(g'',g)\over\bar{p}(g)}.
\label{cofg}\eea

Our calculations replace $G_0$ by $G_0^{(LO)}$ of \eq{glo}.

\section{Simple models and analytic results}\label{sec:models}
One way to verify  the  
\textit{LO} approximation is 
to show that it reproduces  analytic results for previously
published models. We consider the models of \cite{gcalda03} and
\cite{deeds} in this section. 
In both of these models
$p(g,g')$ is taken as a step function (the
0 temperature limit of our model):
\bea p(g,g')= \Theta(g+g'-\mu).\label{sharp}\eea
The two models differ in their choice of $\rho(g)$, but the use of \eq{sharp} allows 
one
to obtain compact general expressions for the  generating functions $\widehat{G}_0(z,g), 
\widehat{G}_0(k,g),p_k $ and $c(k)$. We present these first and discuss specific
details of the individual  models in separate sub-sections.

The use of \eq{sharp} in \eq{gbog} yields
\bea 
\ln \widehat{G}_0(z,g)=
N\left[\Theta(\mu-g)\int_{\mu-g}^\infty dg' \rho(g') +\Theta(g-\mu)\right]\log(z)=N\bar{p}(g)\log(z),
\label{gbogsharp}\eea 
so that
\bea \widehat{G}_0(z,g)=z^{N\bar{p}(g)}.\label{gsharp}\eea 
It is interesting to observe that \eq{gbog1} reduces to the above result.
This is because powers of $p(g,g')^m=p(g,g')$ for \eq{sharp}, so that
the integration appearing in \eq{gbog1} leads to an  expression that is 
a function of $N,z,\bar{p}.$  Then the  use of the binomial theorem allows the
second term of \eq{gbog1} to be expressed as a summable power series in $\bar{p}$
which ultimately leads to the result \eq{gsharp}.

If we follow \cite{bps} 
and treat $k$ as a continuous variable (which requires 
large values of $k$) we find
\bea \widehat{G}_0(k,g)=\delta(k-N\bar{p}(g)) 
=\frac{\delta(g-g_N(k))}{N\left\vert \bar{p}\;'(g_N(k))\right\vert},,\label{gdelta}\\\eea
where $g_N(k)$ is the solution of the equation \bea k=N\bar{p}(g).\label{gn}\eea
Note that  for $k=N$, $g_N(k)$ can take on any value greater than  $\mu$. 
The result \eq{gdelta} is the same as eq.(34) of \cite{bps}, 
but written in a more compact form.
The use of \eq{gdelta} in \eq{ours} and
\eq{eq:21} yields  the results
\bea
p_k={\rho(g_N(k))\over N\left\vert \bar{p}\;'(g_N(k))\right\vert}\\
  \bar{c}(k) ={c(g_N(k))\over N\left\vert \bar{p}\;'(g_N(k))\right\vert}.\label{gdelta1}\eea 

\subsection{Model of Caldarelli {\it et al.}\cite{gcalda03}}
This model is defined by using $\rho(g)=\exp(-g)$, but we generalize
to take the form
\bea \rho_\lambda(g)=\lambda\exp(-\lambda g).\eea
Ref.~\cite{bps} works out this model using their Green's function formalism.  
Our purpose here is
to compare the results of our averaging approximation with their results. 
For this model
 the average interaction probability $\bar{p}(g)$ is given by
\bea
\bar{p}(g)=\int_0^\infty dg'\lambda \exp{[-\lambda g']} \Theta(g+g'-\mu)= \Theta(g-\mu) + \Theta(\mu-g)\exp{[-\lambda(\mu- g)]}.\label{pbarss}\eea
Then our approximation \eq{ours} for  the degree distribution $p_k$  is given by
\bea
p_k=\left(\begin{array}{c}N\\k\end{array}\right)
\int_0^\mu dg \lambda  \exp{[-\lambda g]} \exp{[-k\lambda (\mu-g)]}\left(1- \exp{[-\lambda (\mu-g)]}\right)^{N-k}\eea
Define the integration variable $t\equiv \exp{[-\lambda (\mu-g)]}$  so that
\bea
&&p_{k}=\left(\begin{array}{c}N\\k\end{array}\right)e^{-\lambda\mu}\int_{t_0}^1 {dt\over t^2}t^k(1-t)^{N-k},\quad t_0\equiv e^{-\lambda\mu}\\
&&p_{k>1}=\left(\begin{array}{c}N\\k\end{array}\right)e^{-\lambda\mu}\left({\Gamma(N+1-k)\Gamma(k-1)\over \Gamma(N)}-B_{t_0}(k-1,N+1-k)\right),\label{pkif}\\
&&p_{k=1}=N e^{-\lambda\mu}{(1-t_0)^N\over N}\;_2F_1(1,N;N+1,1-t_0)\eea
where $_2F_1$ is the confluent hypergeometric function 
and $B_{t_0}$ is the incomplete Beta function (and with $t_0=1$  the Beta function):
\bea B_z(a,b)\equiv \int_0^z dt t^{a-1} (1-t)^{b-1}\;,B_1(a,b)=B(a,b).\eea
Consider the case
\bea 1<k,\quad \lambda\mu\approx 10,\eea (the latter is typical of our biological model)
so that the  second term of \eq{pkif} can be neglected. Evaluating the remaining gamma functions gives
\bea
p_k=e^{-\lambda\mu}{N\over k(k-1)}.\label{cutus}\eea
Ref.~\cite{bps} computes the degree distribution for this model 
in analytic manner,  using the approximation \eq{gdelta}
in which $k$ is treated as a continuous variable  and therefore
 ``is expected to perform poorly for small values of $k$''. The result of \cite{bps} $(p_k^{BPS})$ is
\bea p_k^{BPS}=e^{-\lambda\mu}{N\over k^2}+ e^{-\lambda\mu}\delta(k- N)\label{cutum}\eea
which corresponds to agreement (for $k\ne N$) 
within the stated domain of accuracy of Ref.~\cite{bps}. The confluence of
\eq{cutus} and \eq{cutum} provides a verification of the accuracy of the  averaging
 approximation.

The results for $k=N$ seem  to disagree, 
so we examine this more closely.
Use \eq{gsharp} directly to obtain the generating function $G_0(z)$ as 
$G_0(z)=\int dg \rho(g)z^{N\bar{p}(g)}$. One obtains a result $z^N$ for all values of
$g$ ($g>\mu$) such that $\bar{p}(g)=1$.  Using this generating function 
yields the result
\bea p_{k=N}=\int dg\rho(g) \Theta(g-\mu).\eea
The specific value of the integral depends on the choice of $\rho(g)$, but
the result is a finite number  for any choice of $\rho(g)$
that satisfies the normalization condition that its integral over its domain is unity.
Thus
we believe that the correct result of using the propagator (eq(34) of 
\cite{bps} in their eq(11)) is  \bea p_k^{BPS}=e^{-\lambda\mu}{N\over k^2}\label{cutum1}\eea
instead of \eq{cutum}, which is in agreement with our result.

Our approximation works very well in reproducing the  computed clustering coefficient of \cite{bps}.
In particular, we evaluate $c(g)$ of \eq{cofg} to find that
\bea \bar{c}(k)={1\over p_k}\left(\int_0^{\mu/2}\exp(-g)G_0^{(LO)}(k,g)+\int_{\mu/2}^\mu\exp(-g)G_0^{(LO)}(k,g)
(2g-\mu+1)\right).\eea
Numerical evaluation of this approximate expression accurately 
reproduces the result of Fig.~3 of Ref.~\cite{bps}.
Thus our mean field approximation is accurate for both our model\cite{us} and the model of 
Ref.~\cite{gcalda03},

\section{Protein Protein Interaction Network-model of Shi {\it et al.} 
\cite{us}}\label{sec:PIN}
Our principal application is to the 
the PIN of Ref.~\cite{us}.
This model is based on the concept of free energy of association. For a given pair of
proteins the association free energy (in units of $RT$) is assumed to deviate from an average
value a number contributed by both proteins additively as $g+g'$. 
This is a unique approximation
to first-order in $g$ and $g'$. Thermodynamics and
 the assumption that
the interaction probability is independent of concentration allows us to write
\bea p(g,g')={1/ (1+e^{\mu-g-g'})},\label{pdef}\eea which  reduces to a step function in the zero 
 temperature limit, but otherwise
provides a smooth  function. 
Increasing 
the value of $\mu$  weakens  the strength of
interactions, and previous results \cite{us} showed  the existence of an evolutionary trend
 to weaker interactions in
more complex organisms.
 The probability that a protein has a value of $g$ is given by the probability distribution
\bea\rhol={\lambda\over e}e^{-\lambda g}, -1\le\lambda g\le+\infty,\label{rhodef}\eea
where the positive real value of $\lambda$ governs the fluctuations
 of $g$. 
We previously chose the species-dependent values of $\lambda$ and $\mu$ so as to reproduce
measured degree distributions obtained 
using the yeast two-hybrid method (Y2H) that reports binary results 
for protein-protein binding under a controlled setting\cite{fields}. 
Those parameters are displayed in 
Table I. The  impact of the parameters $\lambda$ and $\mu$ are explained in Ref.~\cite{us} and displayed in
Fig.~3 of that reference. Increasing the value of  $\lambda$  increases the causes a more rapid decrease
of $p_k$--the slope of $p_k$ increases in magnitude. Increasing the  value of $\mu$ decreases the
magnitude of $p_k$ without altering the slope  much for values of $k$ greater than about 10.
The ability to  vary both the slope and magnitude of $p_k$ gives this model flexibility that 
allows us to describe the available  degree distributions for different species.

\begin{table}\label{table1}
\caption{Parameters obtained in Ref.~\cite{us}}
\begin{center}
\begin{tabular}{|c|c|c|c|}
\hline
\quad    Species  & N & $\lambda$ & $\mu$ \\
\hline
{\it H. pylori} & 732 &0.88 & 7.06\\
{\it P. falciparum} & 1,310 & 0.93 &7.77\\
{\it S. cerevisiae} &4,386  & 1.18 &7.94\\
{\it C. elegans} &2,800 &1.29&8.19\\
{\it D. melanogaster} &2,806&1.53&8.89\\
Human \cite{rual} &1,494 &0.64 &10.6\\
Human \cite{stelzl} &1,705 &0.67 &10.2\\
\hline
\end{tabular}
\end{center}
\end{table}

\begin{figure}
\unitlength.95cm
\begin{picture}(10,10)(5,0)
\includegraphics[width=10 cm,height=10cm]{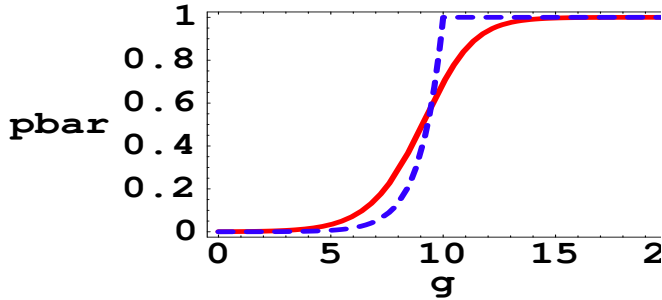}
\end{picture}
\caption{\label{fig:pbar}(Color online)
 Average connection probability $\lambda=1$, $\mu=10$. 
Solid (Red): result of  \eq{special}; dashed (blue) (containing the step function) result of \eq{pbarss}.
The approach to unity is smooth for  \eq{special}.}
\end{figure}

We obtain an analytic form for the for $\bar{p}(g)$ \eq{pave} of this model.
Given \eq{rhodef} and \eq{pdef} we find an analytic result:
 \bea\bar{p}(g,\lambda)=\;_2F_1(1,\lambda;\lambda+1;-\exp{(\mu-g)}),\label{analpbar}\eea
where
$_2F_1$ is the confluent hypergeometric function. The special case $\lambda=1$ yields a closed 
form expression:
 \bea \bar{p}_1(g)=e^{g-\mu}\;\ln(1+e^{\mu-g}).\label{special}\eea
A smooth average connection probability is obtained in contrast
with the result of the sharp cutoff model \eq{pbarss}. This shown in Fig.~\ref{fig:pbar}.

It is useful to define the variable \bea\xi\equiv \exp{(\mu-g)}>0,\eea   
and note that  an integral representation\cite{ab}
\bea _2F_1(n,\lambda;\lambda+1;-\xi)=\lambda\int_0^1\;dt\;t^{\lambda-1}(1+\xi t)^{-n},\label{alg}\eea
is convenient for numerical evaluations.

\begin{figure}
\unitlength.95cm
\begin{picture}(10,10)(2,-1.)
\includegraphics[width=10 cm,height=15cm,angle=90]{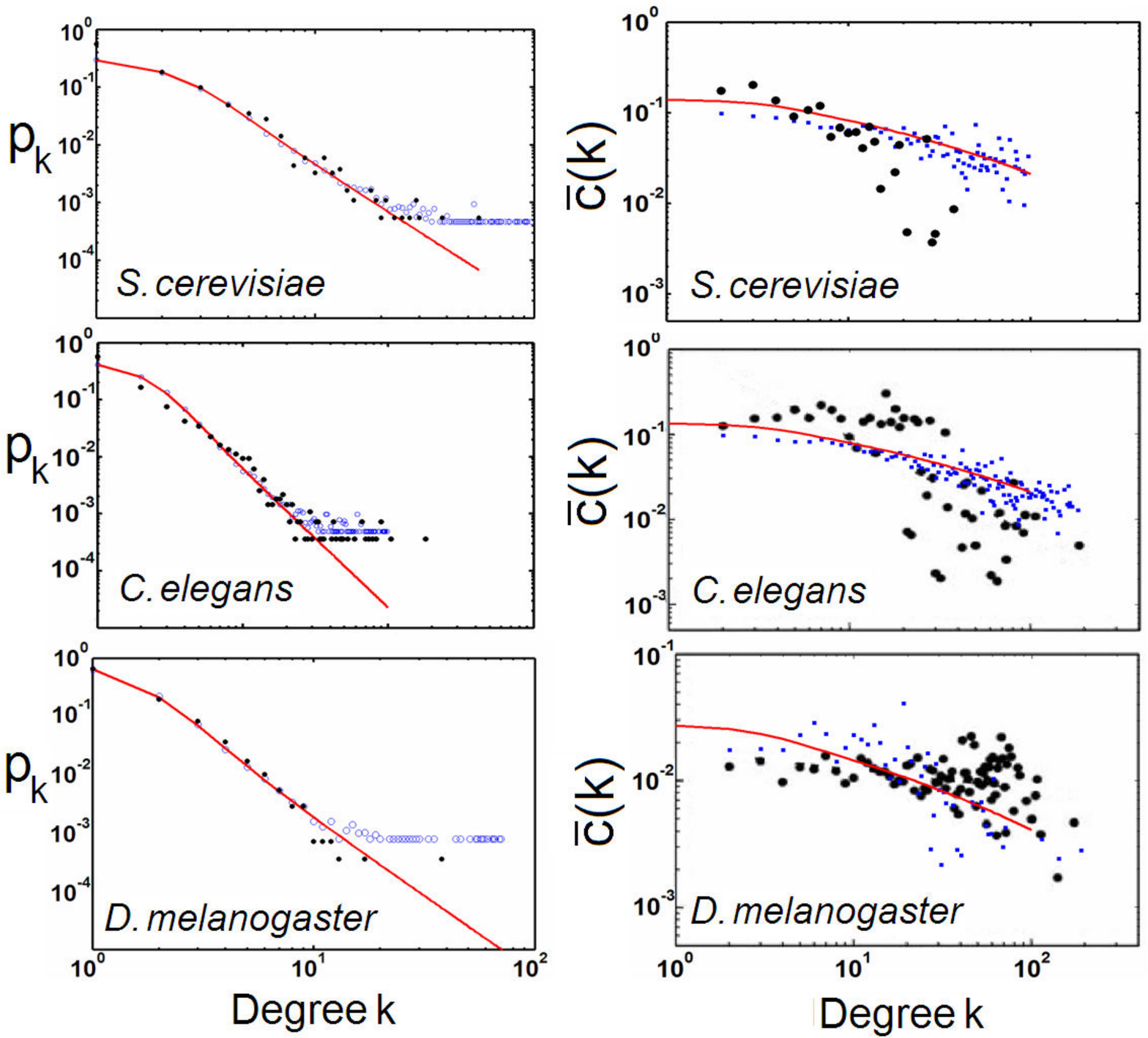}
\end{picture}
\caption{\label{degc}(Color online) Degree distributions $p_k$
and clustering coefficients $\bar{C(k)}$ 
of diverse
species.
 Degree distributions $p_k$:  The solid (red) curves are derived from 
 the $LO$ theory. The black dots are the results of
experimental data as referenced in the text.
The small (blue) circles are the results of 
a numerical simulation using the procedure of  \cite{us}.
Clustering coefficients $\bar{C(k)}$: 
The solid (red) curves are derived from 
 the $LO$ theory. The small (blue) dots are the results of a numerical simulation using the
procedure of \cite{us} and the heavy (black) dots represent  the measured data.
 }
\end{figure}

Knowledge of the propagator \eq{glo} allows  us to compute 
the clustering coefficients of  diverse  species.
The resulting degree distributions
of $p_k$ (shown for the sake of   completeness) and the newly computed 
clustering coefficients $\bar{c}(k)$ for yeast {\it S. cerevisiae} \cite{yeastref}, 
worm {\it C. elegans} \cite{wormref} and  
fruit fly {\it D. melanogaster} \cite{flyref}
  are shown in Fig.~\ref{degc}. The 
parameters $\lambda$ and $\mu$ are those of \cite{us}, so  the calculations of the
clustering coefficients represent an independent major new prediction of our model.
Results of numerical simulations and our analytic procedure are presented.
 The excellent agreement
between the two methods  verifies the $LO$ approximation. 
More importantly,
 the agreement
between our calculations and the measured clustering coefficients is generally very good, 
so our
model survives a very significant test. This bolsters the notion that the properties of a PIN 
are determined by a distribution of free energy.  The clustering coefficient for yeast drops
rapidly for large values of $k$ (where statistics are poor), 
a feature not contained in our model.

It is worthwhile to compare our model with that of \cite{deeds}. 
That work chooses a Gaussian form of $\rho(g)$, based on 
hydrophobicity,  a step function 
form of $p(g,g')$, and is applied only
to yeast. We found \cite{us} that  $p_k$ of  \cite{deeds}
is scale free only for a narrow range of parameters,
and we could not reproduce the data for diverse species using that model.   

\begin{figure}
\unitlength.9cm
\includegraphics[width=9.1 cm]{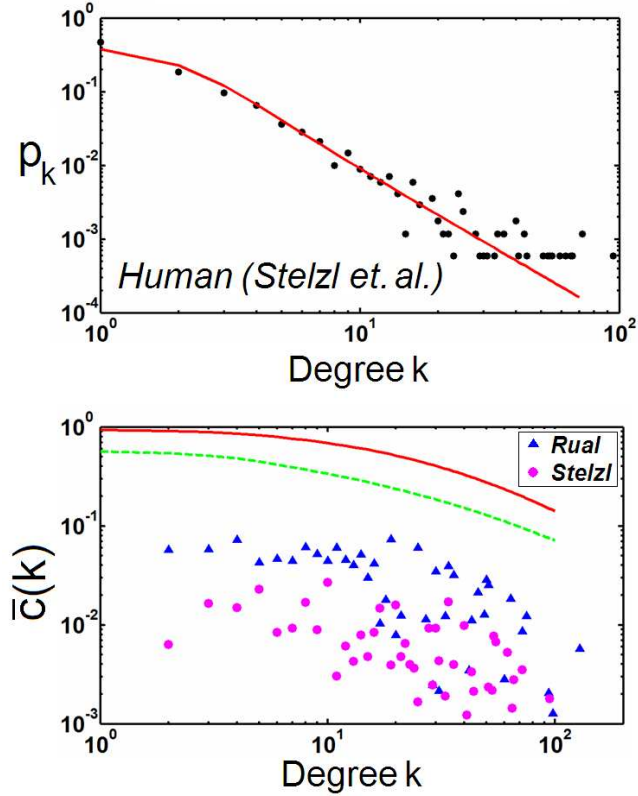}
\caption{\label{hdegc}(Color online) Human degree distribution $p_k$; 
the solid (red)
curve is obtained using both set   A
$\lambda=0.67,
\mu=10.6$ and set B $\lambda=0.94,\mu=8.3$. The black dots represent the experimental data. The data set is that  of \cite{rual}, but nearly identical
data is obtained from \cite{stelzl}.
Human  cluster coefficient $\bar{c}(k)$:
The solid (red) curve  is computed using  set A
$\lambda=0.67,
\mu=10.6$ and the dashed (green) using set B $\lambda=0.94,\mu=8.3$. 
 Measured human clustering coefficients are from 
 \cite{rual} triangles (blue) and \cite{stelzl} heavy dots (pink). 
}
\end{figure}

 The human interactome  is of special interest.
Fig.~\ref{hdegc}A shows the human degree distributions 
computed with two sets of parameters, one from
Ref.~\cite{us} (Table I)  and the other 
using values of $\lambda=0.94, \mu=8.27$ shown in the caption. The degree 
distributions are essentially identical, so only one curve can be  shown. 
Each is approximately of a power law form and each describes the measured degree distribution
very   well\cite{explain}. 
Calculations of degree correlations allows one to distinguish the 
two parameter sets. Figure~\ref{hdegc}B
shows that the cluster coefficients differ by a factor of two. 
We find that $\bar{c}(k)$ decreases substantially
as $\lambda$ 
increases. The increase in  
$\lambda$ reduces the allowed spread in the value of $g$ and reduces the value of
integrand  of \eq{eq:21}. It is interesting to note that
the two  existing measurements of the human $\bar{c}(k)$ 
differ by a factor of about an order of magnitude
 with  the measurements of Ref.\cite{stelzl} 
obtaining much smaller values than those of \cite{rual}. The results
of \cite{rual} are closer to our computed $\bar{c}(k)$
 results for   $\lambda=0.94,\mu=8.3$. In  contrast with
the results for other species, our $\bar{c}(k)$ lie significantly  above the data. 
However, the two data sets disagree substantially (by a factor 
of as much as 100 for certain values of $k$)
and both show a clustering 
coefficient that is generally significantly smaller than that of the other species.
Several possibilities  may account for the  
discrepancies between these two measurements  of $\bar{c}(k)$ in humans  and also for the 
differences between 
our  model predictions and the experimental results.
 i) The human studies sample a limited subset of links 
of the complete network and this could bias the results. 
ii) The human protein subsets used in the two studies differ. 
iii) The human interactome is truly less connected than that of other species. 
This demonstrates the importance of measuring degree correlations to
determine the underlying properties of the network. 
The current model and these considerations suggest the 
need  for better design of future PIN studies  
that will not only include other species, 
but also comparisons between the PINs  of 
different  organs of a given species. Furthermore,  comparisons between
normal and malignant tissues  could also  be very fruitful.

\noindent

\section{Summary and Discussion}
\label{sec:summary}
In summary, 
this work provides a method to obtain the properties of  
 hidden variable network models.
The use of the approximation \eq{pave}, used to obtain the 
propagator \eq{glo},
provides an  excellent 
numerical approximation to exact results for the models considered here. 
If  necessary,
the method can be systematically 
improved through the calculation of higher order corrections.
Our principal example is the PIN of Ref.~\cite{us}. Not only does the use of  \eq{glo} 
provide an accurate  numerical result,
 but the model  correctly predicts the clustering coefficients of most species.
For the human interactome,  two different parameter sets 
yield nearly the same degree distribution 
but very different clustering coefficients, showing 
the importance of measuring degree
correlations to determine the underlying nature  of the network.

This work was supported in part by National Institutes of Health Grants GM45134 and
DK45978 (to K.B.). 
We thank the authors of Refs.~\cite{rual,stelzl}
for providing tables of their data.


\begin{thebibliography}{99}
\bibitem{str} S.~H.~Strogatz, Nature (London) {\bf410},268 (2001).
\bibitem{albert} R.~Albert and A.-L.Barab\'{a}si, Rev. Mod. Phys. {\bf74},47 (2002) 
\bibitem{newmanrev} {M.~E.~J.} {Newman}, SIAM Rev. {\bf45}, 167 (2003).
\bibitem{alexei}
{R.}~{Pastor-Satorras},
A.~V\'{a}zquez,
and
{A.}~{Vespignani},
{Phys. Rev. Lett.} \textbf{{87}},
 {258701} ({2001}).

\bibitem{alexei02}
{{A.}~{V\'{a}zquez}},
 {{R.}~{Pastor-Satorras}},
  {and}
 {{A.}~{Vespignani}},
 {Phys. Rev. E} {\bf 65},
  {066130} ({2002})
\bibitem{nprl}{M.~E.~J.} {Newman},
{Phys. Rev. Lett.} {\bf 89},
  {208701},{2002}.
\bibitem{newmanmixing}{M.~E.~J.} {Newman}, Phys. Rev. E {\bf 67}, 026126 (2003).

\bibitem{cell} B.~Alberts {\it et al.} , {\it The Cell}, (Garland Science, New York 2002).
\bibitem{gcalda03}G. Caldarelli, A. Capocci, 
P.DeLosRios, and M. A. Mu\~{n}oz, Phys. Rev. Lett. {\bf89}, 258702 (2002).
\bibitem{soderberg} B. S\"{o}derberg, Phys. Rev. E {\bf 66}, 066121 (2002).
\bibitem{bps}M.~Bogu\~{n}\'{a}  and R. Pastor-Satorras, Phys. Rev. E {\bf 68}, 036112 (2003).
\bibitem{us} Yi Y. Shi, G.A. Miller, H. Qian, and K. Bomsztyk, Proc. Nat. Acad. Sci. {\bf 103}, 
11527 (2006).
\bibitem{deeds}E.J. Deeds, 
O. Ashenberg, and E.I. Shakhonovich,  Proc. Nat. Acad. Sci. {\bf 103}, 311 (2006).
\bibitem{ab}  M. Abramowitz and I.~A.~Stegun, {\it Handbook of Mathematical Functions}, 
(Dover, New York 1970).
\bibitem{thurner} S.~Abe and S.~Thurner, Phys. Rev. E {\bf 72}, 036102 (2005);
 S.~Abe and S.~Thurner, Int. J. Mod. Phys. C {\bf17}, 1303 (2006). 
\bibitem{fields}S. Fields and S.~Song, Nature {\bf 340}, 245 (1989).
\bibitem{yeastref} http://www.nd.edu/~networks/resources/protein/bo.dat.qz
\bibitem{wormref} S.~Li, {\it et al.}, Science {\bf 303}, 540 (2004).
\bibitem{flyref} L.~Giot {\it et al.},  Science {\bf 302}, 1727 (2003).
\bibitem {explain} The value of $p_0$ is a testable result of our model,
even though experimentalists do not measure this quantity. The predicted number of proteins
with no interactions is $p_0N$, where the value of $N$ is given in Table I.
The experimentalists
conventionally normalize their distributions
as $\sum_{k=1}^\infty\;p_k=1$, so we multiply our 
computed $p_k$ by a factor of $1/(1-p_0)$ so that the computed sum $\sum_{k=1}^\infty p_k$ is unity.
\bibitem{rual}J.F.~ Rual, {\it et al.} Nature {\bf 437}, 1173 (2005)
\bibitem{stelzl}U. Stelzl, {\it el al.} Cell {\bf 122}, 957 (2005)
\end{thebibliography}
\end{document}